\documentclass[aps,pre,twocolumn,showpacs,superscriptaddress]{revtex4-2}

\usepackage{graphicx}  
\usepackage{dcolumn} 
\usepackage{bm}
\usepackage{amssymb} 
\usepackage{amsmath}
\usepackage{dsfont}
\usepackage[english]{babel}
\usepackage{pgfplots}
\usepackage{textcomp}
\usepackage{color}
\usepackage{soul} 
\begin{document}
\widetext

\title{Dried Blood Spot Recovery:\\
A Microfluidic Technique for Fast Elution Without Dilution}

\author{\'Etienne~Coz}
\email[]{etiennecoz@gmail.com}
\affiliation{Microfluidic Laboratory, MEMS \& Nanostructures, from the Institute of Chemistry, Biology and Innovation (CBI) UMR8231, ESPCI Paris, CNRS, PSL Research University, 10 rue Vauquelin, Paris, France}

\author{Alexandre~Vilquin}
\email[]{alexandre.vilquin@espci.fr}
\affiliation{Gulliver UMR 7083 CNRS, PSL Research University, ESPCI Paris, 10 rue Vauquelin, 75005 Paris, France}
\affiliation{IPGG, 6 rue Jean-Calvin, 75005 Paris, France}

\author{\'Elian~Martin}
\affiliation{Microfluidic Laboratory, MEMS \& Nanostructures, from the Institute of Chemistry, Biology and Innovation (CBI) UMR8231, ESPCI Paris, CNRS, PSL Research University, 10 rue Vauquelin, Paris, France}

\author{Pierre~Garneret}
\affiliation{Microfluidic Laboratory, MEMS \& Nanostructures, from the Institute of Chemistry, Biology and Innovation (CBI) UMR8231, ESPCI Paris, CNRS, PSL Research University, 10 rue Vauquelin, Paris, France}

\author{Yannick~Raguel}
\affiliation{Service de biochimie métabolique, hôpital Universitaire Necker Enfants Malades, APHP, Paris, France}

\author{Jean-Fran\c cois~Benoist}
\affiliation{Service de biochimie métabolique, hôpital Universitaire Necker Enfants Malades, APHP, Paris, France}

\author{Fabrice~Monti}
\affiliation{Microfluidic Laboratory, MEMS \& Nanostructures, from the Institute of Chemistry, Biology and Innovation (CBI) UMR8231, ESPCI Paris, CNRS, PSL Research University, 10 rue Vauquelin, Paris, France}

\author{Patrick~Tabeling}
\affiliation{Microfluidic Laboratory, MEMS \& Nanostructures, from the Institute of Chemistry, Biology and Innovation (CBI) UMR8231, ESPCI Paris, CNRS, PSL Research University, 10 rue Vauquelin, Paris, France}

\date{\today}

\begin{abstract}
Dried blood spot (DBS) has risen in popularity due to the ease of sampling, storing, shipping and more. Despite those advantages, recovery of the dried blood in solution for analysis is still a bottleneck as it is manual, time-consuming and leads to high dilutions. To overcome those issues, we have developed a microfluidic chip allowing reversible opening, holding hermetically DBS and forcing the elution buffer through the thickness of DBS. The new technique, validated with clinical samples, is automated, fast, robust, precise, compatible with in-line analysis and leads to a highly concentrated extraction. Moreover, by using model experiments with fluorescein solutions, we show that the elution process is governed by an advection-diffusion coupling commonly known as Taylor dispersion. This new technique could open the way to a new generation of analytical devices to quantify analytes in DBS samples for a wide range of applications.
\end{abstract}

\maketitle

\section{Introduction}
Dried Blood Spot (DBS) is a form of sampling where drops of capillary blood are dried on filter papers. Advantages like the ease of sampling (capillary blood from pin-prick instead of venous blood draw), storage and transport (DBS are mailed), the reduced amount of blood (10-25\,$\mu$L), the stability of analytes and others led to a rapid growth of its use. Applications range from large-scale use (newborn and metabolic screening~\cite{chace2009mass}, pharmacokinetics studies~\cite{spooner2009dried,patel2010facilitating,kissinger2011thinking,beaudette2004discovery}  to discrete use (therapeutic drug monitoring~\cite{coombes1984phenytoin,malm2004automated,meesters2010ultrafast,ter2008quantification}) and use in resources-limited countries (infectious disease management~\cite{snijdewind2012current,anders2012evaluation,chase2012evaluation,hooff2011dried,johannessen2010dried,li2011strategies,rohrman2012paper,sherman2005dried}). Target analyte concentrations are measured by the application of various quantitative analytical techniques such as gas chromatography-mass spectrometry (GC-MS)~\cite{mess2012dried,kong2011evaluation,deng2002gas,spector2007detection}, liquid  chromatography (LC) with different sensors such as MS~\cite{piraud2003esi,la2008rapid,kromdijk2013short,castillo2013tenofovir,ansari2012simplified,nageswara2012lc}, fluorescence~\cite{tawa1989determination,croes1994simple,nageswara2012determination,romsing2011determination,rao2011high}, UV~\cite{green2002high,blessborn2007development,malm2004automated,allanson2007determination}, immunoassay~\cite{la2008rapid,chase2012evaluation,brindle2010serum,la2009new,lin2011combination}, or PCR analysis~\cite{snijdewind2012current,masciotra2012evaluation,vidya2012dried,de2012rapid}.
 
Despite the massive utilization worldwide, some disadvantages remain. First, recovery of the samples leads to high dilution because of the need to cover the DBS with an elution buffer. At least 200\,$\mu$L of the elution buffer is needed per DBS, which is at least 10 times the blood volume on a DBS. The low analyte concentrations obtained induce a need for sensitive and expensive analytical techniques~\cite{meesters2013state,deglon2012direct,edelbroek2009dried,jardi2004usefulness}. This is due to the predominance of concentration-dependent sensors, such as mass spectrometers, used for DBS analysis~\cite{rainville2011microfluidic}. Secondly, only a few automated extraction solutions are affordable to medical analysis departments~\cite{meesters2013state,deglon2012direct,edelbroek2009dried,johnson2011use,blessborn2007development,deglon2009line}. Indeed, most methods classically used are manual, tedious and slow~\cite{arnaud2011technology}. Third, protocols are plentiful:  elution buffers, DBS size, duration of elution, volume of elution vary and depend on the analyte recovered~\cite{snijdewind2012current,edelbroek2009dried,li2010dried,hirtz2015prelevement,mcdade2007drop}. Those issues led to a growing interest in developing a new set of universal techniques allowing miniaturization, automation, high concentration recovery, fast elution, in-line elution of the DBS and others for all possible elution buffers~\cite{demirev2012dried,meesters2013state,manicke2011quantitative,abu2009,liu2011approach,stokes2011determination}.\\
In this context, microfluidic is a promising avenue since it deals with small amounts of liquid. To our knowledge, only one microfluidic method has been proposed to recover DBS~\cite{shih2012dried,jebrail2011digital}. This method, using digital microfluidics, is mainly passive and slow, while only being able to process the recovery of the dried blood with a wetting liquid.

\begin{figure*}[t]
\includegraphics[width = 1\textwidth,trim=0 0 0 1mm,clip=true]{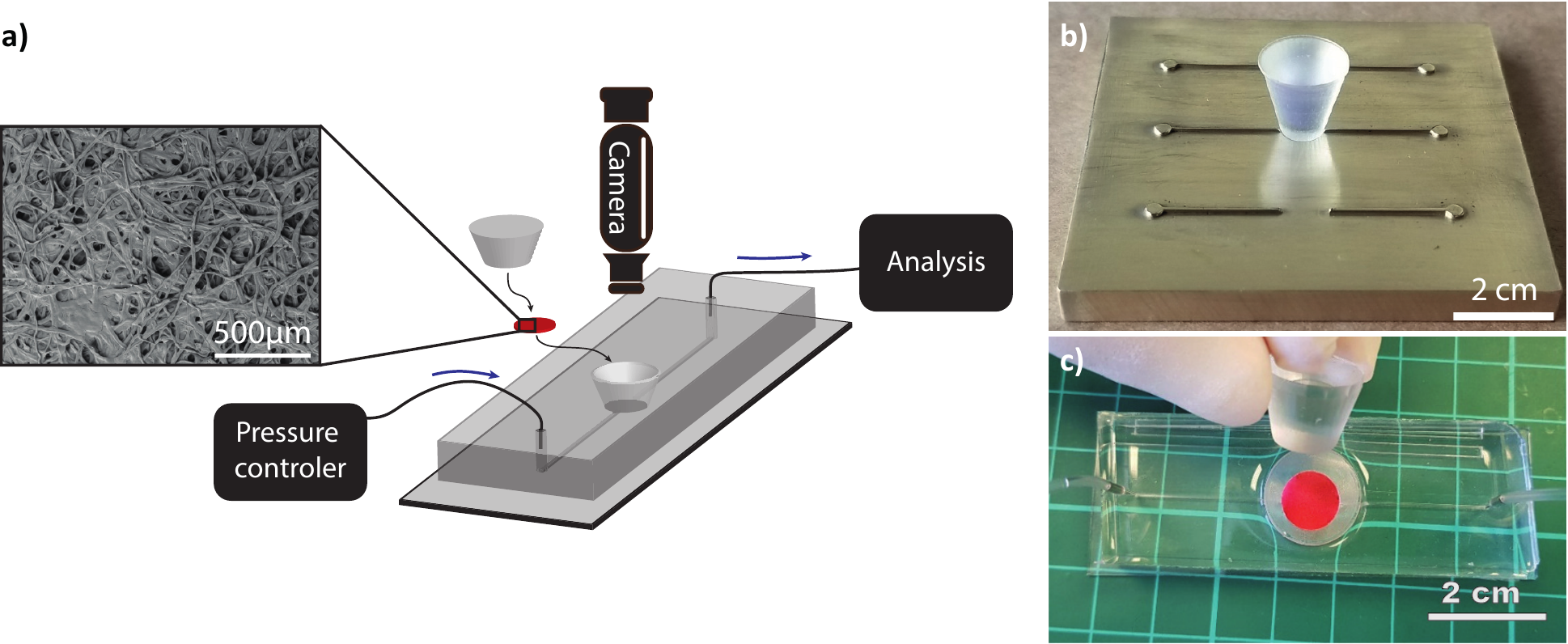}
\caption{a) Schematics and picture of the working principle of the microfluidic elution technology. The microfluidic chip reversibly holds the DBS. A single channel allows the elution buffer to pass through the DBS to recover the dried blood. The enlarged image of the DBS shows an electron microscopy picture of Whatman 903 with dried blood. b) Picture of the brass plate with a 3D-printed truncated cone. c) Picture of the chip ready to be closed with a DBS inside.}
\label{fgr:MicrofluidicChip}
\end{figure*}

We report here both a new process of microfluidic chip conception and a new universal technic for automatic extraction of analytes from DBS. We designed a microfluidic chip that can be opened and tightly closed with a plug that holds a DBS and a single channel allowing a flow of elution buffer to percolate through the DBS. The flow is monitored with a pressure controller. We characterized this new form of microfluidic chips by performing the desorption of a model system, specifically that of the fluorescein from a DBS with a water flow. Our results show a fast recovery of highly concentrated extraction of analytes. This new microfluidic DBS analysis method was also validated by quantifying Phenylalanine (PHE), an amino acid commonly measured as a biomarker for Phenylketonuria, in clinical samples. Phenylketonuria is an inborn error of metabolism, resulting in a decreased metabolism of the amino acid PHE. The associated treatment is a low PHE diet with patients having to monitor their PHE concentration by sending DBS every week to a center of analysis~\cite{BOOK1Phenylketonuria,BOOK2Phenylketonuria}. Finally, we show that our method also works both for wetting and non-wetting elution buffers (here water and methanol), demonstrating the reliability of this new microfluidic technique for the elution with all buffer-DBS interactions.

\section{Experimental}

\subsection{Reagents and Materials}
Unless otherwise specified, reagents were purchased from Sigma Aldrich. Whatman 903 filter papers were purchased from GE Healthcare.

\subsection{Microfluidic Fabrication}
The microfluidic chips were fabricated with polydimethylsiloxane (Momentive RTV615 A) mixed with 10\% cross-linking agent (Momentive RTV615 B) poured on a micro-milled brass plate. As shown in Figure~\ref{fgr:MicrofluidicChip} a) and b), the brass plate surface is made of a single channel (500\,$\mu$m width and height) opened in the center by a free area the size of a classical DBS (3 to 10\,mm). A corresponding double-sided tape disc (3 to 10\,mm diameter and 500\,$\mu$m thickness) of classical DBS was used to stick a 3D-printed truncated cone plug on the brass plate and thus connecting the two sections of the channel. The reticulated elastomers are treated under $O_2$ plasma (Femto Science CUTE) to be covalently bonded to a glass coverslip. A PDMS plug is made using a 3D-printed mold complementary to the 3D-printed plug. The PDMS plug is the exact complementary of the conic hole in the microfluidic chip. A clamping device using 4 screws was used to apply a constant pressure on the DBS allowing sealing of the chip. The plug allows the DBS to be introduced and removed in the PDMS chip. Plugs are made as truncated cones so they are easily introduced to hold the DBS and to seal the chip as shown in Figure~\ref{fgr:MicrofluidicChip} c). This is to our knowledge, the first microfluidic chip allowing a quick and easy access to its inside while staying perfectly hermetic when closed.

\subsection{Microfluidic instrumentation for flow control}
The microfluidic flow is controlled using a pressure controller system (Fluigent MFCS) with an M-flow unit to measure the flow rate. The Fluigent dynamic flow-rate control allows the system to reach a steady state in a few seconds. Automation of the flow control is performed by using the Fluigent script module. Automation is achieved with both target flow rate and final volume. For the elution of DBS with the non-wetting buffer (water), a first high-velocity pulse forces the elution buffer to burst through the DBS. This pulse is created with an instantaneous high pressure command (1000\,mbar), immediately followed by a much lower flow rate, typically 10\,$\mu$L/min for a 5\,mm DBS, up to the desired volume, typically 20\,$\mu$L. 

\begin{figure*}[t]
	\includegraphics[scale = 1.0,trim=0 0 0 1mm,clip=true]{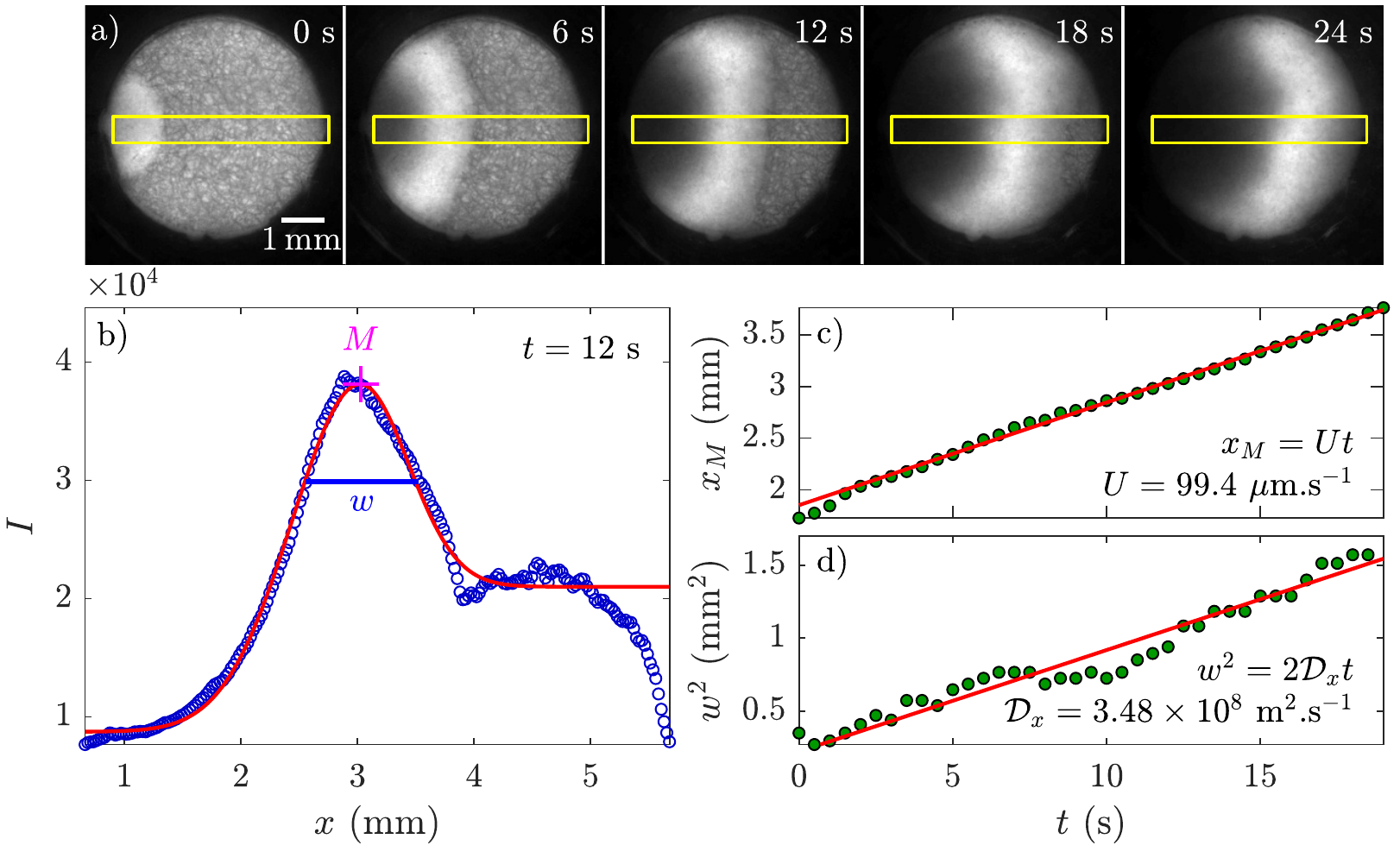}
	\caption{\label{fgr:FrontDynamics} a) Image sequence of the fluorescein elution on a Whatman 903 disc for a flow rate $Q=8$\,$\mu$L.min$^{-1}$. The yellow rectangles show the regions of interest where the intensity profiles are averaged in the transverse direction. b) Example of an intensity profile averaged in a small region shown in a). The red line shows the non-linear fit used to determine both the front position $M$ and width $w$. Temporal evolution of the front position c) and width d). The red lines indicate the linear fits respectively used for the determination of the front velocity $U$ and dispersion coefficient $\mathcal{D}_x$.}
\end{figure*}

\subsection{Dried Blood Spot}
DBS are made using Whatman 903 papers, a class II medical device virtually used in all international newborn screening programs, made from 100\% pure cotton linters with no wet-strength additives~\cite{healthcare2008whatman}. Once spotted, the samples (fluorescein or blood) are allowed to dry for several hours before being stored in a zip bag with desiccants. 3.2, 5 and 10\,mm diameter discs were punched out using biopsy punchers. Once samples are spotted, DBS are used in under a week. Figure~S1 in Supporting Information (SI) shows scanning electron microscope pictures of Whatman 903 paper with and without dried blood. Typical fibers are 20\,$\mu$m wide and up to several millimeters long. Fresh blood spotted on the paper will fill the pores by capillary action before eventually drying. The initial porosity of the DBS is 0.5~\cite{chao2016simultaneous} and the blood is about 80\% water~\cite{beilin1966sodium}. After drying of the blood, the porosity becomes 0.4. Figure~S1 b) confirms that the paper matrix remains porous once the blood is dried, hence allowing the flow of solvent through the DBS pores. 

\subsection{Sample preparation for clinical assay}
Clinical DBS were obtained from phenylketonuria patients. The samples analyzed consist of a single paper disc, 3.2 mm in diameter, punched out from a dried blood spot. PHE was eluted in a 96-well plate by 500\,$\mu$L of methanol added to 20\,$\mu$L of internal standard for 30\,minutes at room temperature with continuous shaking. PHE concentrations were determined using a liquid chromatography tandem mass spectrometry method adapted from \citeauthor{tuchman1999phenylalanine}~\cite{tuchman1999phenylalanine}. Briefly, the separation was achieved on a Kinetecx F5 1.7\,$\mu$m, $2.1\times 100$\,mm column (Phenomenex). 5\,$\mu$L of eluant was introduced directly into the electrospray source of a Xevo TQD (Waters) set in the positive ionization mode. Multiple reaction monitoring (MRM) mode was used for the quantification of the PHE.

\subsection{Fluorescein experiments}
Elution of fluorescein spiked DBS is monitored through the fluorescence intensity in the channel downstream the DBS. Images are recorded with a EM-CCD Hamamatsu ImagEM camera adapted on a Leica Z16 APO microscope with a FAM optic cube. The light source is a Leica EL 6000 triggered by an EG trigger from R\&D Vision. No bleaching of the fluorescein has been observed. A calibration curve to link the fluorescein intensity and concentration was made by using several fluorescein concentrations in solutions (see Fig.~S2 in SI).

\section{Results}

\subsection{Fluorescein elution from DBS}

We performed experiments on the recovery of fluorescein (a fluorescent dye) from DBS with a water flow to characterize the desorption and elution process. As the water flows through the porous disc, the dry fluorescein desorbs from the DBS and we observe a bright circular front propagating on the disc as shown in Fig.~\ref{fgr:FrontDynamics} a) and Movie~S1 (see SI). Moreover, we note an increase of the front width during its propagation. The dynamics of the front broadening can be explained by the so-called Taylor dispersion: When a solute or any microscopic entity is transported by a shear flow with a velocity gradient, the coupling between the Brownian diffusion and streamwise advection leads to an enhanced dispersion as described by Taylor in a seminal work~\cite{Taylor1953}. The solute is first advected by the surrounding flow before a global broadening due to the Brownian motion, which homogenizes the front for very long times as compared to $\tau\sim L^2/D_0$. The duration $\tau$ represents the diffusion time in the transverse direction, with $L$ and $D_0$ respectively the transverse length and diffusion coefficient of the transported entity~\cite{Einstein1905}.

During the elution, the liquid drives the fluorescein, initially on the disc, leading to an intensity peak, followed by a darker area at the front rear due to the fluorescein depletion. To study the dynamics of the front broadening, we only consider a small region centered on the disc (see Fig.~\ref{fgr:FrontDynamics} a)). For each image, the intensity profiles are spatially averaged perpendicularly to the flow as shown in Fig.~\ref{fgr:FrontDynamics} b). Each profile exhibits an intensity peak allowing definition of the front position $x_M$ and width $w$. The temporal evolution of the front position $x_M$ and width $w$ are shown in Figs.~\ref{fgr:FrontDynamics} c) and d). Figure~\ref{fgr:FrontDynamics} c) shows that the front propagates with a constant velocity $U$, which is consistent with the constant flow rate imposed by the pressure controller. Figure~\ref{fgr:FrontDynamics} c) shows that the front width has a quadratic time dependence in agreement with a diffusion regime. The effective diffusion coefficient, also called dispersion coefficient $\mathcal{D}_x$, is very large as compared to the molecular diffusion coefficient expected for the fluorescein \emph{i.e.} $D_0=4.25\times 10^{-10}$ m$^2$.s$^{-1}$~\cite{Culbertson2002}, but consistent with the Taylor dispersion phenomena.

\begin{figure}[t]
	\includegraphics[scale = 1.0,trim=0 0 0 1mm,clip=true]{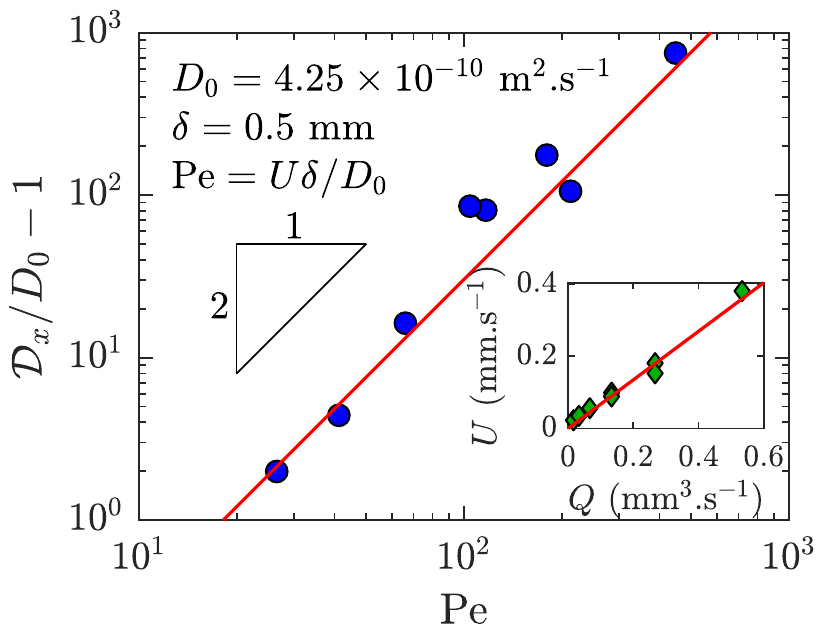}
	\caption{\label{fgr:TaylorDispersion} Dispersion coefficients versus the P\'eclet number. The P\'eclet number is defined as $\mathrm{Pe}=U\delta/D_0$ and the red line indicates a power-law $f\left(x\right)\propto x^2$. The inset shows the corresponding mean velocities $U$ as the function of the imposed flow rates $Q$. The red line indicates a linear regression.}
\end{figure}

Indeed, the dimensionless dispersion coefficient $\mathcal{D}_x/D_0-1$ increases quadratically with the P\'eclet number, defined as $\mathrm{Pe}=U\delta/D_0$ ($\delta$ the disc thickness), as shown in Fig.~\ref{fgr:TaylorDispersion}. This result is in agreement with Taylor's seminal work~\cite{Taylor1953}, where the dimensionless dispersion coefficient obeys to $\mathcal{D}_x/D_0-1=\alpha\mathrm{Pe}^2$ with $\alpha$ a dimensionless coefficient imposed by the channel geometry. Thus, after the desorption, the fluorescein is transported by the surrounding flow and we observe that the dynamics of the front width is governed by the Taylor dispersion for a wide range of flow rates as shown in the inset in Fig.~\ref{fgr:TaylorDispersion}. This enhanced diffusion may facilitate the desorption process of the dry sample due to the faster spreading of the fluorescein in water.

\begin{figure*}[t]
\includegraphics[scale = 1.0,trim=0 0 0 1mm,clip=true]{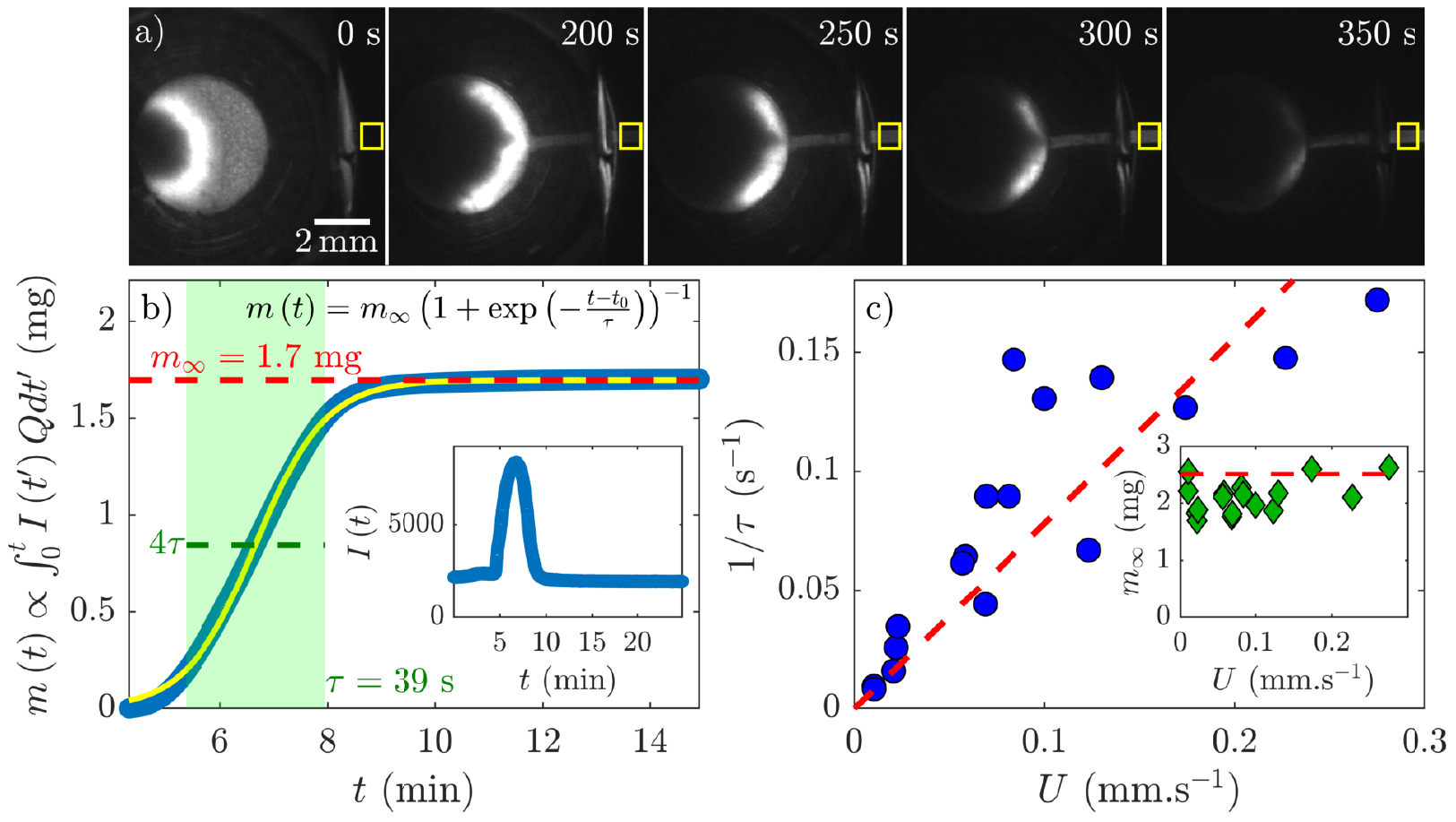}
\caption{a) Image sequence of the fluorescein elution on a Whatman 903 disc for a flow rate $Q=1$\,$\mu$L.min$^{-1}$. The yellow rectangles show the channel where the intensity is averaged. b) Fluorescein mass recovered during the elution process. The yellow and red lines respectively show the hyperbolic tangent fit and the final mass recovered $m_\infty$. The inset shows the intensity profile which is integrated to obtain the recovered fluorescein mass. c) Recovering time $\tau$ as a function of the mean flow velocity $U$. The red line is a linear regression. The inset shows the typical masses recovered $m_\infty$ in all the experiments for an initial deposited mass of 2.5\,mg of fluorescein.}
\label{fgr:FluoresceinElution}
\end{figure*}

We now focus on the recovery of fluorescein extracted by elution from the Whatman disc. We examine the fluorescence intensity peak due to the fluorescein front propagation in the channel downstream as shown in Fig.~\ref{fgr:FluoresceinElution} a). The fluorescein concentration $C\left(t\right)$ is obtained from the intensity signal in the channel (see inset in Fig.~\ref{fgr:FluoresceinElution} b)) by using a calibration curve to link intensity and fluorescence concentration (see Fig.~S2 in SI). Thus, knowing the flow rate $Q$, we integrate the concentration as
\[
 m\left(t\right) = \int_0^t C\left(t'\right)Qdt'\ , 
\]
to obtain the fluorescein mass recovered during the elution process. The recovered mass quickly increases before saturating to a plateau value. Thus it can be described by a hyperbolic tangent function $m\left(t\right)=m_\infty\left(1+\exp\left(-\frac{t-t_0}{\tau}\right)\right)^{-1}$, allowing us to extract both the final recovered fluorescein mass $m_\infty$ and elution duration $\tau$. In Fig.~\ref{fgr:FluoresceinElution} c), we observe that the elution duration decreases with the mean flow velocity $U$ and thus, with the flow rate $Q$, as shown in the inset in Fig.~\ref{fgr:TaylorDispersion}. Moreover, the final recovered fluorescein mass $m_\infty$ does not vary with the flow rate.

We show that the recovered fluorescein mass can be captured by a hyperbolic tangent function. As the increase of this function occurs in approximately $4\tau$ (see Fig.~\ref{fgr:FluoresceinElution} b)), $t_0 = n\tau$ with $n\sim 2-3$ if we assume that the fluorescein desorbs in the first instants. As respectively shown in Fig.~\ref{fgr:FluoresceinElution} c) and the inset in Fig.~\ref{fgr:TaylorDispersion}, the elution duration $\tau$ and mean flow velocity $U$ can be expressed as $U=D/\tau$ and the flow rate $Q=SU$, where the distance $D$ and surface $S$ both depend on the microfluidic system. Consequently, the elution duration is given by $\tau=SD/Q$ and by using the dimensionless time $T=t/\tau=Qt/\left(SD\right)$, while the recovered mass can be expressed as
\[
 m\left(T\right)=\frac{m_\infty}{1+\exp\left(-\left(T-n\right)\right)}\ .
\]
Finally, as the volume of the recovered solution is defined by $V=Qt$, the concentration of the solution extracted by elution is given as a function of the dimensionless time $T$. Indeed, $C=m/V$ and it becomes
\[
 C\left(T\right)=\frac{C_\infty}{\left(1+\exp\left(-\left(T-n\right)\right)\right)T}\ ,
\]
with $C_\infty=m_\infty/\left(SD\right)$. Interestingly, for the range of parameters used in our experiments ($SD\approx2.4$\,mm$^3$ and $n=3$) this function can be non-monotonic. Thus, the flow rate $Q$ can be wisely selected, with respect to the parameters of the microfluidic system ($S$ and $D$) to optimize the concentration of the recovered solution.

\subsection{Elution of clinical samples with methanol: a wetting-buffer specifically recovering dried blood's amino acids.}

\begin{figure*}[t]
\includegraphics[scale = 1.0,trim=0 0 0 1mm,clip=true]{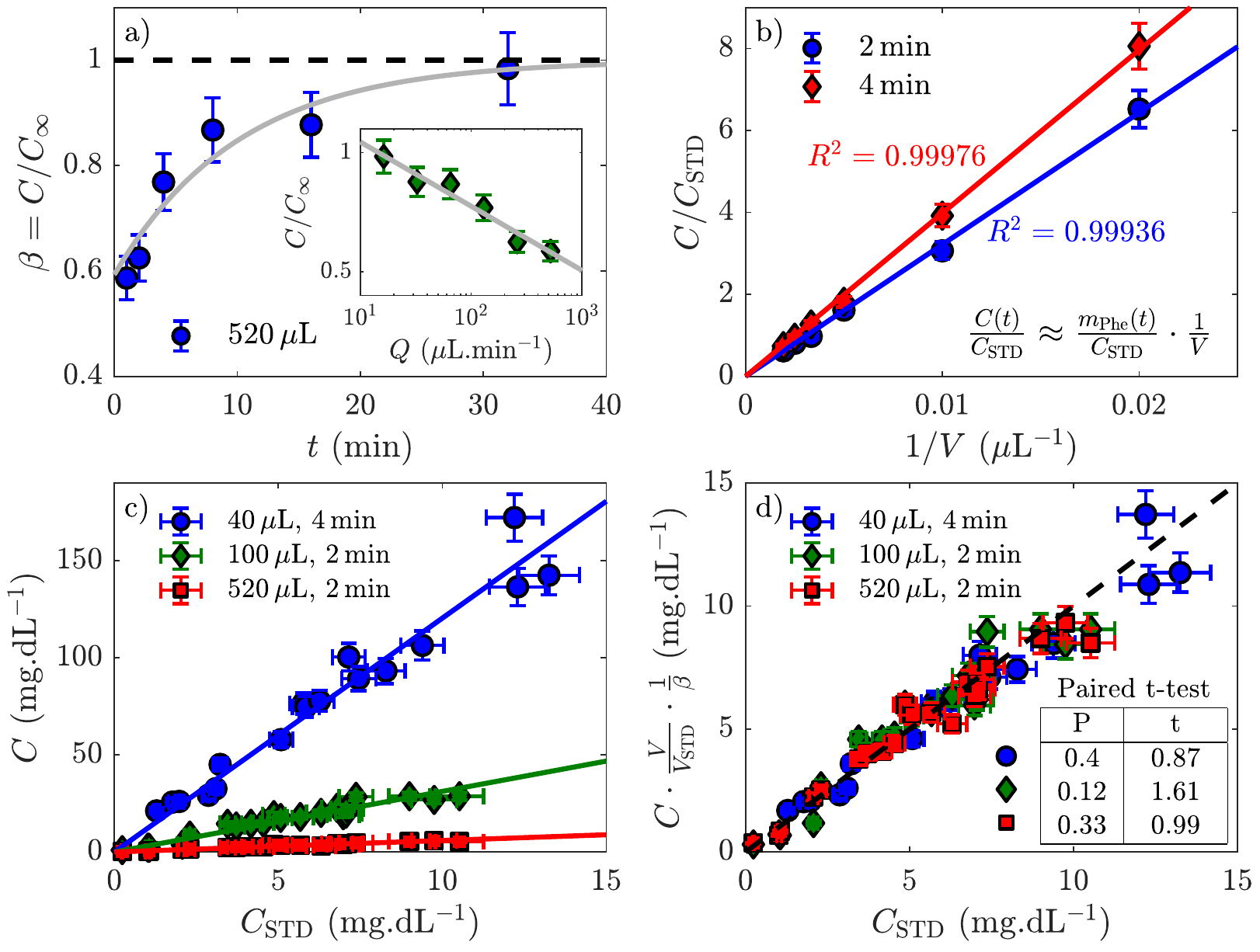}
\caption{Evaluation of the microfluidic technique on various clinical samples for the non-wetting buffer methanol. a) Recovery factor $\beta$ as a function of the elution duration. DBS are from a single patient. b) PHE concentration as a function of the volume V of elution, for two different times of elution. DBS are from a single patient. c)  Raw data for the elutions of DBS from various patients under 3 sets of elution volume and time $V$ and $T$. d) Normalized data of the previous graph c) with paired t-test results in insert.}
\label{fgr:ElutionMeOH}
\end{figure*}

The microfluidic method previously described is now used on clinical samples from phenylketonuria patients to quantify the concentration of the amino acid Phenylalanine (PHE). By contrast with our microfluidic technique, the standard technique is a 30-minute passive elution in wells with 520\,$\mu$L of (25:1) methanol: isotope-labeled standard. Methanol instantaneously precipitates the blood proteins while amino acids are eluted. Figure~\ref{fgr:ElutionMeOH} a) shows the recovery ratio of PHE $C(t)/C_\infty$ as a function of time $t$ for an elution with a volume $V=520$\,$\mu$L with the microfluidic technique for clinical sample. The time $t$ is varied over one order of magnitude by changing the flow rate $Q$. The concentration $C(t, V=520\,\mu L)$ increases with $t$ and tends to level off after 20 minutes. After 4 minutes, we have a valuable recovery factor $\beta=C/C_\infty=75\%$, where $C_\infty$ is the concentration value in long-time limit. After 30 minutes, the concentration of PHE recovered by the microfluidic technique is equal to that of the standard technique. As shown in the inset of Figure~\ref{fgr:ElutionMeOH} a), the flow rates are larger than those used for the fluorescein experiments, which may explain the smaller values of the recovery factor.

Figure~\ref{fgr:ElutionMeOH} b) shows the concentration $C$ of PHE obtained with the microfluidic technique, normalized by the concentration $C_\mathrm{STD}$ obtained with the standard method, as a function of the volume $V$ for two elution times $t$. The volume $V$ is varied over one decade by changing the flow rate. First, we note that the concentration ratio becomes quickly larger than 1. Moreover, it increases linearly as a function of $1/V$, demonstrating that the PHE mass $m$ recovered by the microfluidic technique is only a function of the elution time as $C\left(t\right)=m\left(t\right)/V$. Hence, a small elution volume leads to high concentration of the analyte recovered. This way to obtain larger concentrations has been tested for several clinical samples with various PHE masses in the DBS. Fig.~\ref{fgr:ElutionMeOH} c) shows the concentration $C$ retrieved with our microfluidic method as a function of the concentration $C_\mathrm{STD}$ retrieved with the standard method. The concentration of the clinical samples recovered with the microfluidic technique is up to 10 times higher than the one recovered with the standard technique. Fig.~\ref{fgr:ElutionMeOH} d) shows the same data as in Fig.~\ref{fgr:ElutionMeOH} c) once normalized by both volumes and recovery factors $\beta$. A paired t-test shows no significant statistical difference between the two methods at a 95\% confidence level. P and t values of the paired t-test are shown in the table of Fig.~\ref{fgr:ElutionMeOH} d). Finally, the recovery factors allow to predict the larger concentration values that can be obtained with our microfluidic technique.

\subsection{Elution of clinical samples with water: a non-wetting buffer recovering all the dried blood analytes.}

DBS are used in many situations and the protocols for retrieving back the dried blood in solution are plentiful. However, we can sort the elution buffers by their affinity with the dried blood on the paper matrix, whether the elution buffer wets the DBS or not. Wettability of the dried blood depends on the elution buffer: While methanol easily wets the DBS, water does not (see Fig.~S3 in SI). In the microfluidic chip, percolation of a non-wetting fluid through the DBS tends to create large and randomly distributed dry areas. These drainage issues have already been studied by \citeauthor{lenormand1990liquids}~\cite{lenormand1990liquids}: Three main regimes exist for the displacement of a wetting fluid (air in our case) by a non-wetting fluid (water in our case). These three behaviors depend on two dimensionless numbers: the capillary number and viscosity ratio $R$, respectively defined as $\mathrm{Ca}=U\eta_\mathrm{injected}/\gamma$ and $R=\eta_\mathrm{injected}/\eta_\mathrm{initial}$, with $U$ the streamwise mean velocity of the moving fluid, $\eta_\mathrm{injected}$ and $\eta_\mathrm{initial}$ the viscosities of the moving and quiescent fluids, $\gamma$ the surface tension. As the two fluids here are liquid and gas (water and air in the example), the ratio of viscosity $R$ is always constant and larger than 1 in our experiments. Thus, only two of the three possible regimes are attainable: capillary fingering for low capillary numbers and stable displacement for large capillary numbers. They correspond to the limit when two of the three forces involved during displacement can be neglected. Capillary fingering occurs when the injection rate is low and viscous forces are negligible in both fluids. Stable displacement occurs at large flow rate when capillarity forces are low with a negligible pressure drop in the displaced phase.

\begin{figure}[t]
\includegraphics[scale = 1.0, trim = 0 0 0 0]{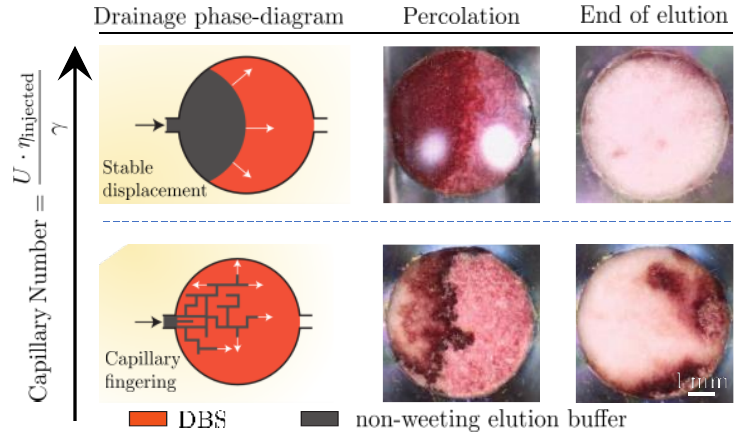}
\caption{Drainage of liquid in a porous media. Viscous forces, capillary forces and pressure drop are responsible for the fluid percolation pattern. Three main regimes exist. Two are accessible for high viscosity ratio ($\eta_\mathrm{water} >> \eta_\mathrm{Air}$). Stable displacement at high rate when capillary forces are low and pressure drop is negligible in the displaced phase while capillary fingering occurs when capillary forces are dominant. Photos of the two possible regimes for DBS elution are shown for percolation and after elution. DBS are 5 millimeter in diameter. Percolation flow rate is $Q=10$\,$\mu$L.min$^{-1}$ for the capillary fingering example. Percolation pressure is 1000\,mbar for the stable displacement example. Elution flow rate is $Q=10$\,$\mu$L.min$^{-1}$ and elution volume is 20\,$\mu$L for the two examples.}
\label{fgr:LiquidInPorousMedia}
\end{figure}

Figure~\ref{fgr:LiquidInPorousMedia} shows the phase-diagram for drainage for our system, adapted from \citeauthor{lenormand1990liquids}~\cite{lenormand1990liquids}.
Capillary fingering is observed for a typical flow rate of $Q=10$\,$\mu$L.min$^{-1}$. Some areas remain dry when capillary fingering occurs during percolation of DBS by a non-wetting buffer. Those area will eventually disappear as the elution buffer dissolves the dried blood from the interface. This time depends on the dried surface area but is substantially longer than the time needed to dissolve dried blood from a properly wet area. Furthermore, the percolation stops when a capillary finger reaches the end of the DBS and connects to the microfluidic channel. Consequently, large randomly distributed areas remain dry. In order to achieve elution under two minutes without dilution, DBS percolation has to follow the stable displacement regime. The value of the capillary number can only be modified by changing the velocity of the injected fluid. To fully and homogeneously wet the DBS, we apply an initial substantial velocity pulse, with an instantaneous high-pressure command (1000\,mbar), thus increasing the capillary number, before imposing a lower flow rate to complete the elution without dilution.

\begin{figure}[t]
\includegraphics[scale = 1.0,trim=0 0 0 1mm,clip=true]{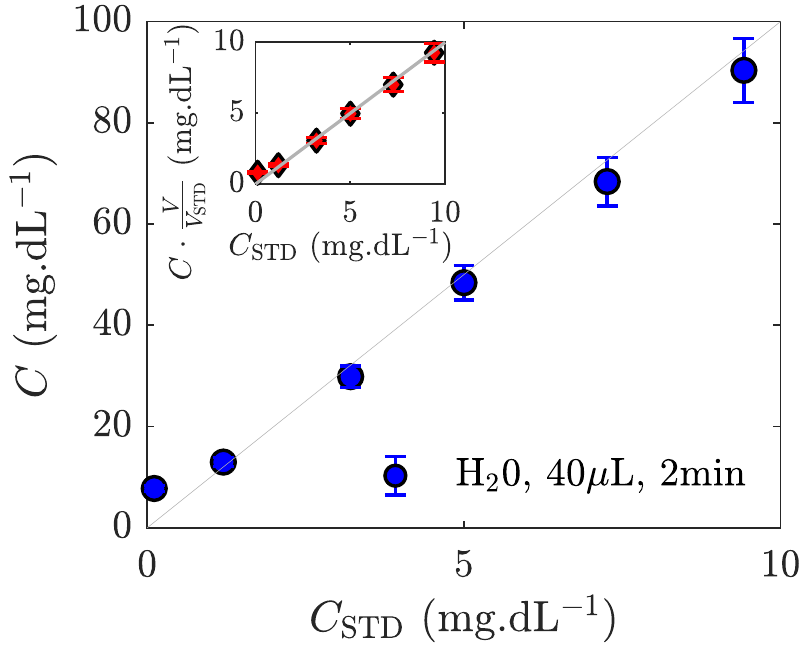}
\caption{Evaluation of the microfluidic technique on various clinical samples for the non-wetting buffer water. Elutions are 2 minutes long and elution volumes are 40\,$\mu$L.} 
\label{fgr:ElutionH2O}
\end{figure}

To validate our microfluidic technique with all possible dried blood-elution buffer interactions, we have evaluated the elution of clinical samples from phenylketonuria patients with water: a non-wetting buffer recovering all the dried blood back in solution. The standard technique is an overnight passive elution in wells with 520\,$\mu$L of (25:1) water: isotope-labeled standard. Our microfluidic protocol is automatized. A sharp high-velocity pulse homogeneously wets the dried blood spot and is followed by a 2-minute low flow rate leading to 40\,$\mu$L elution volume to avoid the capillary fingering, as previously described. Figure~\ref{fgr:ElutionH2O} shows the concentration $C$ of the samples retrieved with the microfluidic technique as a function of the concentration $C_\mathrm{STD}$ of the same DBS retrieved with the standard technique.
The concentration of the clinical samples recovered with the microfluidic technique is up to 10 times higher than that recovered with the standard technique. The insert in Fig.~\ref{fgr:ElutionH2O} shows the same data once normalized by volumes. A paired t-test shows no significant statistical difference between the two methods at a 95\% confidence level. This example demonstrates the reliability of our microfluidic method for various situations.

\section{Conclusion}
For elution of dried blood spots, we have developed an automated microfluidic device to elute of dried blood spot. The dried blood spot is hermetically and reversibly held in the microfluidic chip. The elution buffer is driven through the DBS.\\
Firstly, we performed experiments on the recovery of a fluorescent dye as a model system, highlighting a Taylor dispersion phenomenon in elution process. We demonstrate that the final recovered mass of fluorescein does not depend on the flow rate, thus the flow rate can be set in order to recover a highly concentrated solution. Here, the elution is fast (under 2 minutes) and leads to the recovery of fluorescein in solution at large concentrations as compared to the standard methods.\\
Secondly, we characterized the different possible interaction between the elution buffer and the dried blood on the DBS. Certain buffers, such as methanol, easily wet the DBS whereas some, such as water, do not wet the DBS. In the case of the elution of a DBS with a non-wetting buffer, a drainage issue may appear, creating randomly distributed dry areas on the DBS and leading to incomplete elution. A protocol has been defined to wet the DBS with a stable displacement regime of percolation by using an initial high-pressure pulse.\\
Finally, for both wetting and non-wetting elution buffers, we demonstrate the utility of our method on clinical samples to measure the phenylalanine concentration, an amino acid commonly used as a biomarker for Phenylketonuria. We obtained as much as 10 times more concentrated solutions than the standard method, with a duration approximately 10 times shorter. We showed with a paired t-test that the normalized results both in volume of elution and recovery factors showed no significant statistical differences.\\

To conclude, this new protocol can allow elution under two minutes of dried blood spot, leading to a highly concentrated sample while allowing the use of less sensitive analytical devices.

\section*{Outline}
This study was performed according to the French Public health regulations (Code de la Santé Publique - Article L1121-3, amended by Law n°2011-2012, December 29, 2011 - Article 5) using anonymized residues of DBS sampled for the follow-up of phenylketonuria patients. No other biological investigation was performed other than PHE analysis.

\medskip

\begin{acknowledgments}
\noindent The authors thank Aditya Jha and Emer Buckley for patient reading. This work has benefited from the technical contribution of the joint service unit CNRS UAR 3750. The authors also benefited from the financial support of CNRS, the Fondation ESPCI, the Institut Pierre-Gilles de Gennes (Equipex ANR-10-EQPX-34 and Labex ANR-10-LABX- 31), and PSL Research University (Idex ANR-10-IDEX-0001-02).
\end{acknowledgments}

\section*{Supporting Information Available}

The following files are available free of charge.
\begin{itemize}
	\item MovieS1.mp4: The Movie S1 shows the elution process, with the front propagation and broadening, in the fluorescein experiment described in Figure 2. The parameters are thus detailed in the caption of this figure.
	\item supp.pdf: The supporting information file contains additional Figures S1, S2, S3, respectively related to the microscopic structure of the Whatman 903 paper disc with and without dried blood, the calibration curve relating image intensity and fluorescein concentration, interaction between DBS and two elution buffers. This last figure is accompanied by a short discussion on the effect of wetting and non-wetting buffers on the elution process.
\end{itemize}


%

\end{document}